# Improvement of superconducting properties of $FeSe_{0.5}Te_{0.5}$ single crystals by Mn substitution

A. Günther[1], J. Deisenhofer[1], Ch. Kant[1], H.-A. Krug von Nidda[1],
V. Tsurkan[1,2*], and A. Loidl[1]

[1]*Experimental Physics 5, Center for Electronic Correlations and Magnetism,
Institute of Physics, University of Augsburg, D-86159, Augsburg, Germany*
[2]*Institute of Applied Physics, Academy of Sciences of Moldova, MD-2028, Chisinau,
R. Moldova*

*Abstract.* We report on structural, susceptibility, conductivity, and heat-capacity studies of $FeSe_{0.5}Te_{0.5}$ single crystals with 2% substitution of Mn for Fe. Mn-doped samples show a higher onset temperature, a narrower width of the superconducting transition, and a higher magnitude of the jump in the specific heat at $T_c$ in compared to undoped samples. From the resistivity data in magnetic fields parallel to the *c* axis we derived an upper critical field $H_{c2}$ of ~580 kOe for doped samples compared to 490 kOe for pure samples. Using a single-band BCS model we can describe the electronic specific heat in the superconducting state with a gap $\Delta(T=0) = 31$ K for the Mn-doped sample in comparison to 26 K for pure $FeSe_{0.5}Te_{0.5}$.

**I. Introduction**

Superconductivity in iron-based pnictides [1-3] and chalcogenides [4] is a hot topic in solid-state and materials science. The iron chalcogenides forming the so-called "11" group are believed to require a much simpler description than pnictides with a more complicated structural arrangement. Slightly off-stoichiometric FeSe exhibits superconductivity at relatively low temperatures (~8 K) [4], however, the critical temperature $T_c$ can be enhanced by external pressure up to 37 K [5, 6]. The superconducting properties of FeSe depend critically on the stoichiometry [7-9]. They also can be changed by different substitutions on the cation and anion sites. For example, substitution of Fe by transition metals such as Ti, V, Co, Ni, and Cr destroys superconductivity [10, 11]. The substitution of Se by Te in FeSe increases $T_c$ up to ~14 K for 50% of replacement, e.g., $FeSe_{0.5}Te_{0.5}$ [12, 13], but $T_c$ is suppressed with further increase of the Te concentration. Moreover, substitution of S for Te also induces superconductivity in FeTe and enhances the amount of the superconducting phase in FeSe [11, 14, 15]. In most cases mentioned above, bulk superconductivity is difficult to achieve. Indeed, in the best explored "11" system, $FeSe_{1-x}Te_x$, bulk superconductivity is reported only for x~0.5, whereas for other concentrations the superconductivity is only filamentary. Even for the composition x=0.5 the volume fraction of the bulk superconducting phase and the width of the superconducting transition vary rather significantly depending on details of the preparation route [12, 16-20]. At present the origin of this behavior is far from being understood. The extreme sensitivity of the properties of the iron chalcogenides to minor deviations from the stoichiometry makes the elaboration of methods to stabilize their superconducting properties highly necessary.

Here we report on the properties of the superconducting FeSe$_{0.5}$Te$_{0.5}$ single crystals with substitution of 2% Fe by Mn ions as studied by magnetic susceptibility, resistivity, and specific heat. We find a clear increase of the onset temperature, a narrowing of the superconducting transition, and an increased magnitude of the jump in the specific heat at $T_c$ in the Mn- doped samples compared to those for the pure samples. Besides that, the doped samples exhibit a lower value of the susceptibility in the normal state indicating a smaller content of magnetic impurities.

## 2. Experimental

Single crystals of pure and Mn-doped FeSe$_{0.5}$Te$_{0.5}$ were grown by self-flux method in identical conditions. As the starting materials we used high-purity elements, 99.99 % Fe (chips), 99.999 % Se (chips), 99.999 % Te, and 99.99% Mn powder. To reduce the amount of oxide impurities, which have a significant influence on the superconducting properties [21], we additionally purified Se and Te by zone melting. Handling of the samples was performed in an argon box with residual oxygen and water content less than 1 ppm. Single crystals were grown in double quartz ampoules sealed under vacuum of $10^{-4}$ mbar. Initial treatment was performed at 650 $^{o}$C for 10 h followed by heating to 700 $^{o}$C for 24 h. Further heating was performed up to 1100 $^{o}$C with 72 h soaking at this temperature. After this the ampoule was cooled with a rate of 1 $^{o}$C/min down to 400 $^{o}$C for final annealing during 100 h followed by quenching in ice water. The composition of the samples was checked by Energy dispersive x-ray analysis (EDX). The EDX data are reported elsewhere [21]. The phase content of the samples was also analyzed by x-ray powder diffraction (Cu K$_\alpha$ radiation, $\lambda$ = 1.540560 Å) on crushed single crystals using a STADI-P powder diffractometer (STOE & CIE) with a position sensitive detector.

Magnetic measurements were performed in a temperature range 2 - 400 K and in magnetic fields up to 50 kOe using a SQUID magnetometer MPMS 5 (Quantum Design). The heat capacity was measured by relaxation method using a Quantum Design physical properties measurement system (PPMS) in a temperature range 1.8-300 K and magnetic fields up to 90 kOe. Resistivity studies were performed on rectangular samples by four-point method using the resistivity-measurement option of the PPMS with electrical contacts made of silver paint.

For comparison, we also show the data for the best prepared undoped sample (labeled as F216 step 1 in [21]).

## 3. Experimental results and discussion

The x-ray diffraction pattern for the Mn-doped sample together with the refined spectrum using the FULLPROF SUITE [22] is shown in Fig. 1. The x-ray data were refined within tetragonal symmetry P4/nmm [23] for the main FeSe$_{0.5}$Te$_{0.5}$ phase and within hexagonal symmetry P63/mmc for the $Fe_7Se_8$ impurity phase. No other impurity phases were revealed by x-ray diffraction. The positions of Se and Te at the 2c sites were refined with different z coordinates. The occupation of Te and Se was refined constraining the sum to unity in correspondence with the EDX analysis. A similar constraint was used for the occupation of Fe and Mn ions in the main phase. For the Fe ions two different sites (2a and 2c) [24] were allowed. The occupation factor for Mn was fixed at a nominal level of 2%. The results of the refinement for pure and doped samples are given in Table 1. Within the accuracy of the refinement we could not resolve the exact position of Mn. However, an enhanced value of the lattice constant $c$ compared to the undoped samples suggests that the Mn ions occupy the 2c sites. If the larger Mn ions occupy the 2a positions an increase of the $a(b)$ parameter will be expected, while the experimental data exhibit an opposite trend. Therefore we concluded that the Mn ions preferably occupy the 2c sites. The refined occupation factors for Se and Te are close to their nominal concentration. The refinement reveals a small amount of Fe ions (5%) present at the 2c sites in accord with observations in pure samples [21]. The amount of the hexagonal impurity phase of 4.7% found in the doped samples was higher than in the pure samples (1.4%). Rather astonishingly, the width of the reflections for the doped sample was narrower than for the pure sample (see inset in Fig. 1).

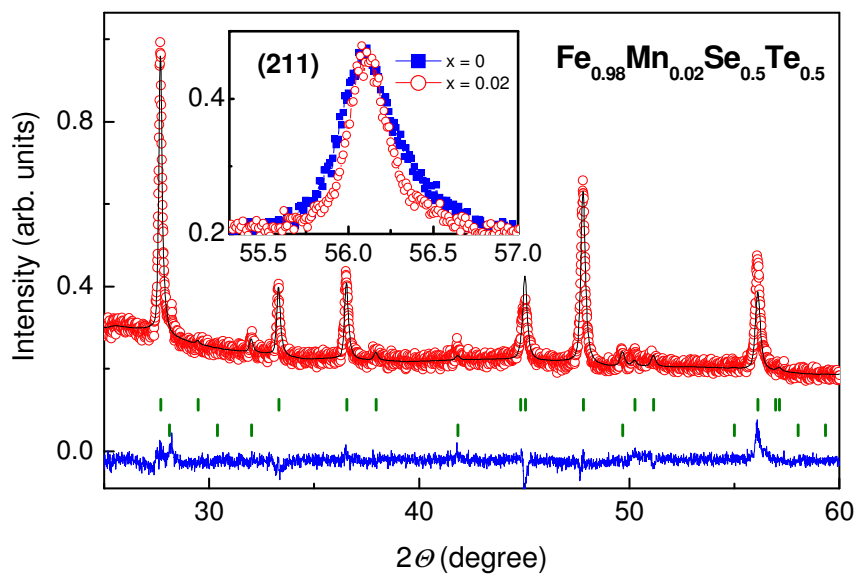

**Figure 1.** (color online) Experimental (red open circles) and refined (black solid line) x-ray diffraction patterns of $Fe_{0.98}Mn_{0.02}Se_{0.5}Te_{0.5}$. The bottom (blue) line represents the difference between the experimental and calculated intensities. The vertical (green) bars mark the Bragg positions of the main tetragonal phase (top) and of the impurity hexagonal phase (bottom). Inset shows the (211) reflection for the pure (blue circles) and doped (red squares) samples.

**Table 1.** Structural data obtained from the Rietveld refinement of $Fe_{1-x}Mn_xSe_{0.5}Te_{0.5}$

| Mn content, x | Occupation Fe1 /2a/ | Occupation Mn2 /2c/ | Occupation Fe2 /2c/ | Occupation Se /2c/ | Occupation Te /2c/ | Lattice constant $a, b$ [Å] | Lattice constant $c$ [Å] | Tetragonal phase [%] | Hexagonal phase [%] |
|---|---|---|---|---|---|---|---|---|---|
| 0 | 0.929(3) | - | 0.071(3) | 0.49(1) | 0.51(1) | 3.8025(3) | 6.0300(9) | 98.6 | 1.4 |
| 0.02 | 0.931(4) | 0.02 fixed | 0.049(4) | 0.50(2) | 0.50(2) | 3.8013(3) | 6.0600(9) | 95.3 | 4.7 |

Fig. 2a shows the temperature dependence of the zero-field cooled (ZFC) and field-cooled (FC) susceptibility for the doped sample measured in a field of 10 Oe applied along the *c* axis. The ZFC susceptibility evidences a sharp transition into the superconducting state with an onset temperature $T_c^{on}$ of 14.4 K which is higher than for the pure sample (13.9 K). The transition width, determined as the difference between the onset temperature and the intercept of the steepest part of the susceptibility extrapolated to the temperature axis, is markedly smaller for the Mn-doped samples (1.0 K) than for the undoped sample (1.5 K). The value of the FC susceptibility is rather low indicating strong flux-pinning. The diamagnetic ZFC susceptibility is by more than two orders higher that the FC susceptibility. The calculated value of $4\pi\chi$ from the ZFC data at 2 K is far above unity suggesting an influence of demagnetizing effects. Measurements of needle-like samples cut from the original samples with a negligible demagnetizing factor in magnetic fields applied along the long axis yielded a value of $4\pi\chi$ close to unity suggesting bulk character of the susceptibility.

Fig. 2b shows the temperature dependences of the magnetic susceptibility measured on cooling in a field of 10 kOe along the *c* axis in the extended temperature range 2 K < *T* < 400 K. The susceptibility of the doped sample manifests non-monotonous temperature dependence with a broad maximum at around 180 K, similar to result observed in the undoped sample. However, the overall variations of the susceptibility for the doped sample are much more pronounced in the normal and in the superconducting states. Beside this, the doped sample exhibits a lower susceptibility in the normal state. Previous studies of $FeSe_{0.5}Te_{0.5}$ single crystals prepared under different conditions [21] have shown that iron oxide (magnetite, $Fe_3O_4$) is the main magnetic impurity present in samples handled in air or prepared from non-purified elements. The susceptibility of the samples containing oxide impurities is significantly higher than that of the oxygen-free samples. The pure sample has the minimal content of the magnetic oxide impurity [21]. Therefore, even a smaller value of the magnetic susceptibility of the Mn-doped sample may indicate a further reduction of magnetic impurities. We also must note that the doped sample contains a nearly three times higher amount of the impurity phase of $Fe_7Se_8$ than the pure sample (Table 1). This suggests that $Fe_7Se_8$ has an insignificant effect on the magnetic and superconducting properties of the doped samples and confirms the earlier conclusion of Ref. 21 which excluded $Fe_7Se_8$ from factors suppressing bulk superconductivity in $FeSe_{0.5}Te_{0.5}$. It also must be mentioned that the larger drop of the susceptibility at $T_c$ and the absence of any upward behavior towards the lowest temperatures as observed in the doped sample suggests a more robust superconducting state resulting from Mn substitution.

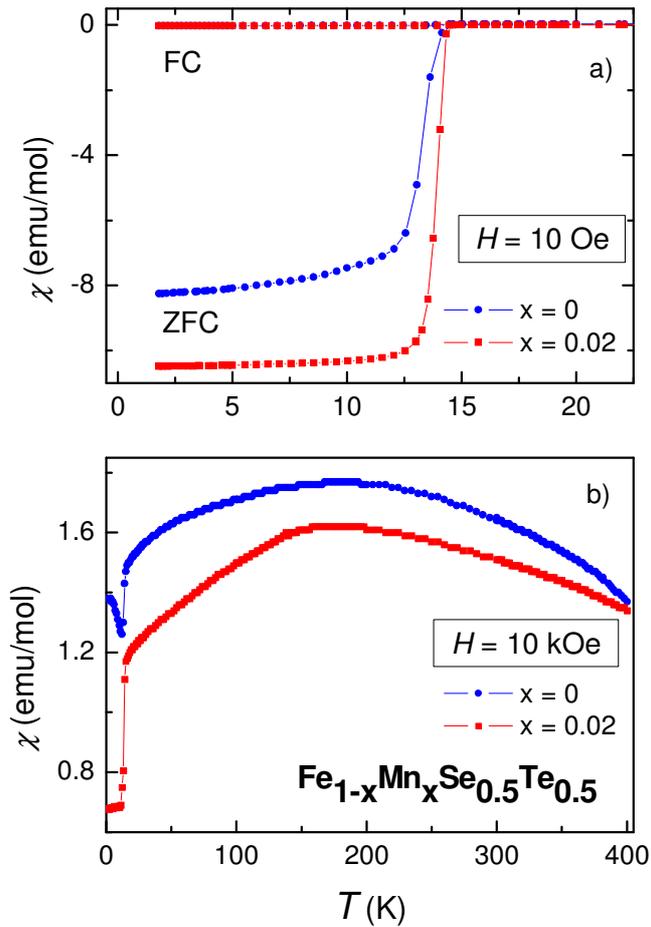

**Figure 2**. (color online): a) Temperature dependences of ZFC and FC susceptibilities for pure (x=0) and doped (x=0.02) single crystalline samples measured in a field of 10 Oe applied along the $c$ axis in the low-temperature range. b) Temperature dependences of the susceptibility for the samples measured in a field of 10 kOe applied along the $c$ axis in the range 2 K < $T$ < 400 K.

Fig. 3 presents the left half (a) of a symmetric magnetization hysteresis loop for the Mn-doped sample measured at different temperatures. In the same figure the 2 K data for the pure sample are shown by open circles. The field dependence of the critical current density $j_c$ estimated using the Bean model for hard superconductors [25, 26] is shown in the right half (b) of Fig. 3 for different temperatures. At 2 K the critical current density $j_c$ of $8.5 \times 10^4$ A/cm$^2$ at zero field was determined for the Mn-doped sample. For the undoped samples we obtained a similar value of $j_c$ at zero field. At the same time, at higher fields the doped samples exhibit larger critical currents (by ~20%) than the undoped ones indicating the presence of additional pinning centers. Above 20 kOe up to the largest measured fields the critical current is only slightly field dependent suggesting high current-carrying ability. The inset of Fig. 3b compares the temperature dependences of the critical current $j_c$ at $H = 0$ for doped and pure samples. For the pure sample the value of the critical currents $j(0) = 1.7 \times 10^5$ A/cm$^2$ (for $T=0$ K) was estimated from the fit to the experimental data using a power-law dependence $j(T) = j(0)[1-(T/T_c)^p]^n$, with p = 0.5, n = 1.5 and $T_c$ = 13.8 K. For the doped sample such an extrapolation was not possible in the whole measured temperature range, but from the experimental data of this sample one can expect a similar high value of the critical current density for $T = 0$ K. We additionally notice that $j_c$ in the doped sample decreases with temperature not as fast as in the pure sample, indicating a higher current-carrying ability on approaching $T_c$. The critical current density calculated from the hysteresis loops at 2 K together with the critical temperature $T_c$ determined from the magnetic data are given in Table 2.

Fig. 4a presents the temperature dependences of the resistivity for the doped sample at around the superconducting transition compared with the data for the pure sample. Similarly to the susceptibility data, the resistive transition for the doped sample is significantly steeper and is shifted by 0.5 K to higher temperatures. The onset temperature of the superconducting transition for the doped sample is at 14.9 K. The resistivity of the doped sample reveals a metal-like increase above $T_c$ up to 200 K similarly to that observed in the pure samples [21]. Such a metal-like behavior was established earlier for FeSe$_{1-x}$Te$_x$ with a low amount of excess iron [27]. In the normal state the Mn-doped sample exhibits a higher resistivity compared to the pure sample which suggests an increased scattering of charge carriers on impurity centers which can be associated with the Mn ions.

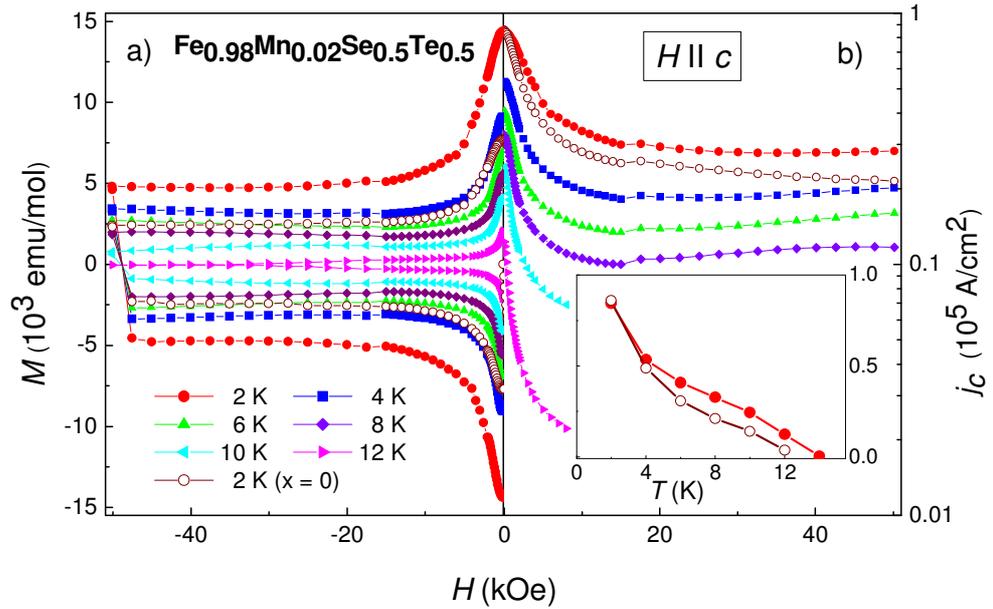

**Figure 3.** (color online) a) Hysteresis loops of the Mn-doped sample (full symbols) at different temperatures and of the pure sample (open circles) for 2 K measured with the field applied along the $c$ axis. b) Critical current density $j_c$ vs. magnetic field at different temperatures for the Mn-doped sample (full symbols) and for the pure sample at 2 K (open circles). The inset shows the temperature dependence of the critical currents at zero field for pure (open circles) and doped (closed circles) samples. Sample dimensions: pure - $1.65 * 3.2 * 0.5$ mm$^3$; doped - $3.4 * 4.75 * 0.5$ mm$^3$.

In Fig. 4b the temperature dependences of the resistivity taken at different magnetic fields in the transition range are presented for the Mn-doped sample. The magnetic field was applied parallel to the $c$ axis. The measurements were performed on warming after cooling in zero field. The resistivity curves exhibit a gradual shift to lower temperatures with increasing magnetic field, similar to report on pure samples [21]. The temperature dependences of the upper critical field $H_{c2}(T)$ determined using the criterion of 50 % drop of the normal-state resistivity $R_n$ is presented in the inset of Fig. 4b. The values of the upper critical field $H_{c2}(0)$ for $T=0$ K were estimated using the expression $H_{c2}(0)=-0.69T_c(dH_{c2}(T)/dT)|_{T_c}$ defined by the Werthamer-Helfand-Hohenberg (WHH) model [28]. The calculated results are presented in Table 2. The estimated value of $H_{c2}(0) \sim 580$ kOe is higher for the Mn-doped samples than for the pure sample (490 kOe) and can be probably attributed to enhanced impurity scattering from the Mn ions.

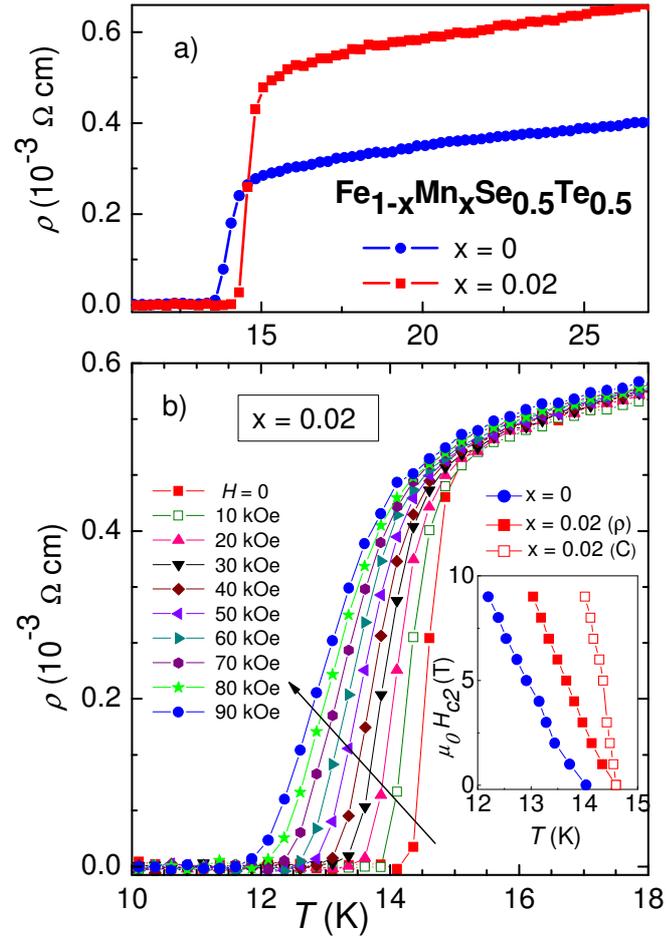

**Figure 4**. (color online). a) Temperature dependences of the in-plane resistivity for pure and Mn-doped samples measured in zero field. b) Temperature dependences of the in-plane resistivity for the Mn-doped sample measured at various magnetic fields applied parallel to the *c* axis. The inset shows the temperature dependences of the upper critical field $H_{c2}$ for pure and Mn-doped samples (closed symbols) calculated from the resistivity data. Open squares represent $H_{c2}(T)$ for the doped sample calculated from the specific heat and shifted by 0.52 K on the temperature axis to fit the onset of the resistivity data.

Fig. 5 shows the temperature dependence of the specific heat *C* for the Mn-doped sample. Above the superconducting transition up to a temperature of 150 K, *C*(*T*) it is very close to that of the pure sample shown in the same figure. The upper inset in Fig. 5 illustrates the specific heat in the low-temperature range. A pronounced anomaly at $T_c$ is evidenced with a much larger specific-heat jump for the Mn-doped sample when compared to the pure one. Magnetic field (applied parallel to the *c* axis) suppresses the anomaly in the specific heat at $T_c$ displacing it to lower temperatures (see inset in Fig. 6). From the shift of the minimum of the temperature derivative of the specific heat in the transition region the upper critical field $H_{c2}$ was determined. It is shown by open squares in the inset of Fig. 4. We found a significant difference between the $H_{c2}(T)$ determined from the resistivity and from the specific heat, the later being much closer to $H_{c2}(T)$ derived from the resistivity curves for the field parallel to the *ab*-plane [21]. The estimations using the WHH formula [28] gave a value of $H_{c2}(0)$ ~1650 kOe which is by a factor of 2.8 larger than that obtained from the resistivity data. A similar large difference in $H_{c2}(0)$ determined from the resistivity and specific heat was reported recently for Ba(K)Fe$_2$As$_2$ [29] and FeSe$_{1-x}$Te$_x$ (with x=0.52) [30] and was ascribed to an anisotropic vortex dynamics.

In the lower inset in Fig. 5 the temperature dependences of the specific heat for the doped sample are shown as $C/T$ vs. $T^2$ at temperatures below 4.5 K, measured in zero field and in a field of 90 kOe. By a fit to the experimental data in the range below 4.5 K using the expression $C/T = \gamma + \beta T^2$ we determined the values of the residual Sommerfeld coefficient $\gamma_r$, related to electronic contribution, and the prefactor $\beta$ characterizing lattice contribution to the specific heat. The respective data are given in Table 2. For the Mn-doped sample a value of $\gamma_r$ ~1.9 mJ/mol K$^2$ was obtained which is two times larger than that of the pure sample (~1 mJ/mol K$^2$ [21]). These extremely low values of $\gamma_r$ are to the best of our knowledge the smallest reported so far for FeSe$_{1-x}$Te$_x$ and, thus, confirm the high quality of our samples.

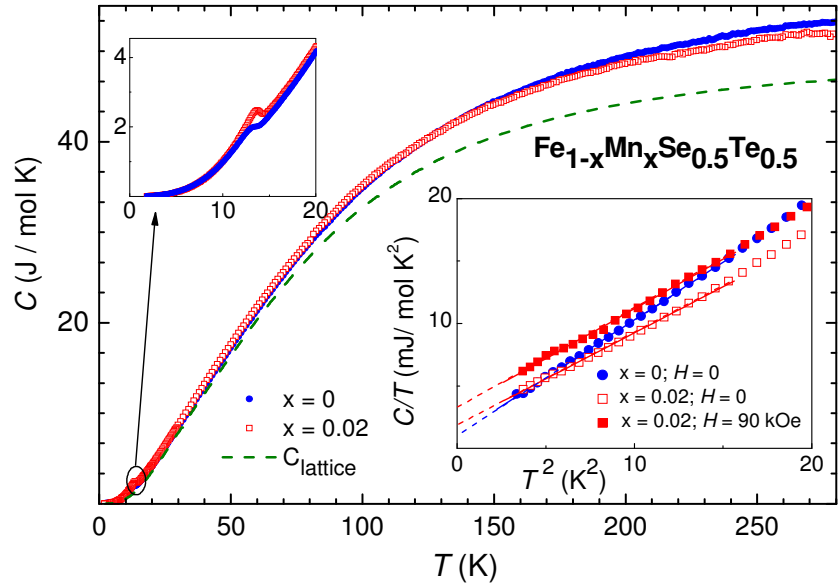

**Figure 5**. (color online) Temperature dependences of the specific heat for pure and Mn-doped samples. The dashed line represents the lattice specific heat calculated using two Debye terms and one Einstein mode with the values of $\Theta_{D1}$ = 127 K, $\Theta_{D2}$ = 235 K and $\Theta_E$ =315 K. The upper inset shows the specific heat in the transition region on an enlarged temperature scale. The lower inset shows the temperature dependences of the specific heat in the representation $C/T$ vs. $T^2$ for the temperature range 1.8 - 4.5 K for $H$=0 and 90 kOe.

The dependences of the electronic specific heat in the representation $C_e/T$ vs. $T$ are shown in Fig. 6 for a temperature range around the superconducting transition. The electronic specific heat was calculated by subtraction of the lattice contribution from the total specific heat. For the calculation of the phonon contribution a combined Einstein-Debye model was used. The justification of this procedure is presented in Ref. 21. The temperature dependence of the calculated lattice specific heat for the doped sample is shown by a dashed line in Fig. 5.

The temperature behavior of the electronic specific heat in the superconducting state was analyzed within the BCS derived α-model [31, 32] with a temperature dependent superconducting gap Δ similar to the analysis of the specific heat in related (Ba:K)Fe$_2$As$_2$ pnictides [33, 34]. The fit results for the electronic specific heat are shown in Fig. 6 by solid lines.

For the ratio of the residual Sommerfeld coefficient $\gamma_r$ to that of the normal state $\gamma_n$ for the Mn-doped sample we obtained a value ~0.07 corresponding to a volume fraction of the superconducting phase of 93%. This is slightly lower than ~96% obtained for the pure sample. Despite the lower volume fraction of the superconducting phase, the doped sample manifests a significantly higher magnitude of the jump in the specific heat at the superconducting transition. Certainly this fact has to be attributed to the substitution effect. It may result, for example, from the increased density of states at the Fermi level, as can be concluded from the higher value of the normal Sommerfeld coefficient (Table 2). Note that the amount of the non-superconducting phase in both samples roughly correlates with the amount of Fe$_7$Se$_8$. However, this impurity phase, as was already noted above, does not suppress the superconductivity of the FeSe$_{0.5}$Te$_{0.5}$. Therefore in samples prepared without Fe$_7$Se$_8$ impurity one would expect a lower residual Sommerfeld coefficient.

We found that the single-band BCS fit reasonably describes the superconducting specific heat, except the range below 5 K which can be probably related to effects of residual impurities. The value of the superconducting gap at 0 K is determined as $\Delta_0$ = 31 K for Mn-doped sample and is higher than $\Delta_0$ = 26 K obtained for the pure sample [21]. This value is in good agreement with the value of 29 K obtained by Kato *et al*. [35] from the tunneling spectroscopy and with the low-energy gap observed by Homes *et al*. [36] in the optical conductivity of FeSe$_{0.45}$Te$_{0.55}$. An enhanced value of the coupling constant $2\Delta_0/T_c$ = 4.47 derived for the Mn-doped sample compared to the pure sample (3.57) exceeds the BCS value of 3.53.

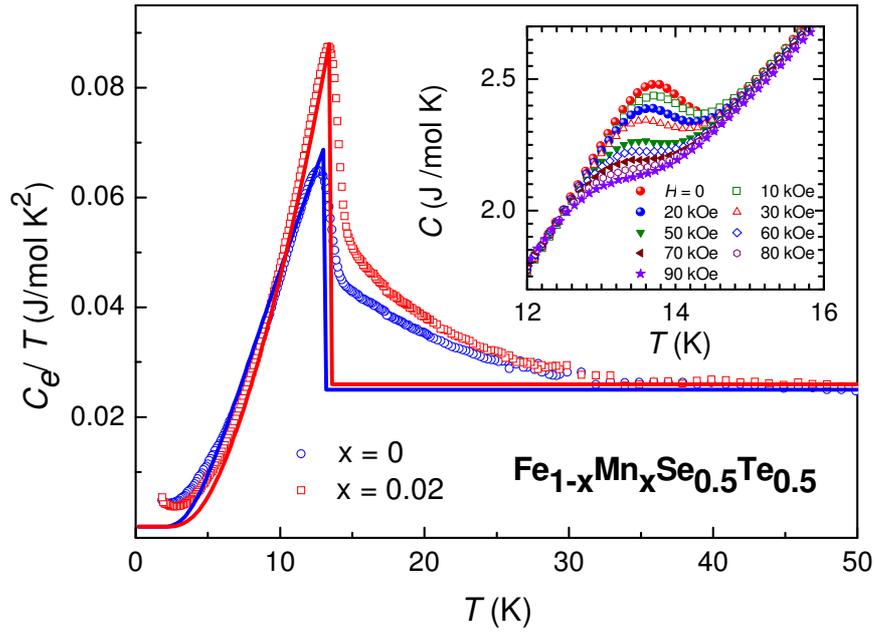

**Figure 6.** (color online) Temperature dependences of the electronic specific heat as $C_e/T$ vs. $T$ for Mn-doped (open squares) and pure (open circles) samples. The solid lines represent the fits describing the superconducting specific heat within the BCS model. The inset shows the temperature dependences of the specific heat in different applied magnetic fields for the Mn-doped sample.

**Table 2.** Parameters of superconducting and normal state for $Fe_{1-x}Mn_xSe_{0.5}Te_{0.5}$ calculated from the magnetic, resistivity and specific heat data.

| Mn content, x | $T_c^{onset}$ (K) from $\chi_{ZFC}$ | $T_c^{ons}$ (K) from $\rho$ | $j_c$ (2 K) (kA/cm$^2$) | $H_{c2}(0)$ (kOe) [$H \parallel c$] | $\gamma_r$ (mJ/mol K$^2$) | $\beta$ (mJ/mol K$^4$) | $\gamma_n$ (mJ/mol K$^2$) | $\Delta_0$ (K) | $2\Delta_0/T_c$ |
|---|---|---|---|---|---|---|---|---|---|
| 0 | 13.9 | 14.4 | 86 | 490* 1500** | 0.96 | 0.94 | 25 | 25.9 | 3.57 |
| 0.02 | 14.4 | 14.9 | 85 | 580* 1650** | 1.88 | 0.74 | 26 | 31.1 | 4.47 |

*Estimated from the resistivity; ** estimated from the specific heat

## 4. Concluding remarks

In conclusion, our studies of the properties of $FeSe_{0.5}Te_{0.5}$ single crystals doped with 2% Mn reveal a clear change of their structural, magnetic and superconducting parameters. The doped samples show narrower x-ray diffraction lines than undoped samples suggesting a higher homogeneity. The lower value of the susceptibility in the normal state for the doped samples indicates a smaller content of the magnetic impurities compared to the undoped samples. The Mn doping obviously has a positive effect on the superconducting properties. Although the observed increase of the onset temperature $T_c^{on}$ (by ~0.5 K) for the doped sample is not large, we observed a pronounced narrowing of the superconducting transition and an enhanced magnitude of the jump in the specific heat at $T_c$ compared to the undoped samples. For the doped samples the critical current density at high fields and the upper critical field are also notably enhanced compared to those for the undoped samples.

We note that very recently an enhanced $T_c^{on}$ of 14.9 K in the resistive transition on polycrystalline FeSe$_{0.5}$Te$_{0.5}$ doped with 5% Mn was reported by Zhang *et al.* [37] supporting our single-crystalline data. However, there is a notable difference between their susceptibility data and our results. Beside this, their data for the resistivity in the normal state show an opposite trend compared to our data.

Considering the enhancement of the superconducting parameters in FeSe$_{0.5}$Te$_{0.5}$ by Mn doping, it should be noted that this effect is in striking contrast to "122" Fe-based superconducting systems, where Mn doping leads to pair-breaking resulting in a considerable reduction of the superconducting transition temperature even at such a small level of substitution as 2% [37]. The mechanisms of pair-breaking in Fe-based superconductors are far from being established. Our present experiments allow to exclude several reasons for the observed changes of properties of FeSe$_{0.5}$Te$_{0.5}$ by Mn doping. They confirm the previous conclusions [21] about the insignificant role of the hexagonal impurity phase of Fe$_7$Se$_8$ in suppressing the superconductivity in FeSe$_{0.5}$Te$_{0.5}$. Discussing the role of the magnetic Fe ions at the 2c site which are assumed to suppress the superconductivity in the 11 system [14, 20, 24, 27], we would like to note that studies of the undoped FeSe$_{0.5}$Te$_{0.5}$ samples with a high volume fraction of the superconducting phase and of those with strongly suppressed superconductivity put a question mark on the validity of this assumption, at least for the samples with a low concentration of the excess Fe [21]. To our opinion, the most plausible explanation of the observed doping effect is related to a reduction of the residual ferrimagnetic iron oxide impurities due to formation of antiferromagnetic manganese oxides. As was already established in Ref. 21, samples containing magnetic oxide impurities exhibit an enhanced susceptibility in the normal state, a reduced onset temperature and reduced magnitude of the jump of the specific heat at $T_c$ compared to samples with a lower content of oxide impurities. The amount of the residual iron oxide impurities in the best pure samples discussed in [21] is below 0.1% as estimated from the change of their susceptibility compared to the impure samples. A further reduction of the susceptibility, an increase of the transition temperature $T_c$, and strong enhancement of the jump in the specific heat at $T_c$ observed in the Mn-doped sample suggests that the significant changes of the materials properties are caused by a rather subtle variation of tuning parameters, most probably, due to residual iron oxide impurities. Of course, for a larger concentration, the substitution can have an opposite effect and Mn can behave in a similar way as the other transition metals that suppress the superconductivity in FeSe [10, 11] and $FeSe_{0.5}Te_{0.5}$ [37]. It is clear that to clarify the role of doping and the origin of the observed changes of the magnetic and superconducting parameters, complete doping series are necessary. These experiments are currently in progress. However, already the present results demonstrate a substantial effect of Mn doping on the properties of $FeSe_{0.5}Te_{0.5}$ and we hope that they will stimulate further experimental and theoretical studies of the interesting "11" superconductors.

## Acknowledgments

The authors thank Dana Vieweg for experimental support. This research has been supported by the DFG via SPP 1458 and Transregional Collaborative Research Center TRR 80 (Augsburg Munich).

## References

[1] Kamihara Y, Watanabe T, Hirano M and Hosono H 2008 *J. Am. Chem. Soc.* **130** 3296
[2] Rotter M, Tegel M and Johrendt D 2008 *Phys. Rev. Lett.* **101** 107006
[3] Tapp J H, Tang Z, Lv B, Sasmal K, Lorenz B, Chu C W and Guloy A M 2009 *Phys. Rev.* B **78** 060505.
[4] Hsu F C, Luo J Y, Yeh K W, Chen T K, Huang T W, Wu P M, Lee Y C, Huang Y L, Chu Y Y, Yan D C, Wu M K 2008 *Proc. Natl. Acad. Sci. U.S.A.* **105** 14262
[5] Medvedev S, McQueen T M, Troyan I A, Palasyuk T, Eremets M I, Cava R J, Naghavi S, Casper F, Ksenofontov V, Wortmann G and Felser C 2009 *Nature Materials* **8** 630
[6] Margadonna S, Takabayashi Y, Ohishi Y, Mizuguchi Y, Takano Y, Kagayama T, Nakagawa T, Takata M and Prassides K 2009 *Phys. Rev.* B **80** 064506
[7] McQueen T M, Huang Q, Ksenofontov V, Felser C, Xu Q, Zandbergen H, Hor YS, Allred J, Williams A J, Qu D, Checkelsky J, Ong N P and Cava R J 2009 *Phys. Rev.* B **79**, 014522
[8] Pomjakushina E, Conder K, Pomjakushin V, Bendele M and Khasanov R 2009 *Phys. Rev.* B **80** 024571
[9] de Souza M, Haghighirad A A, Tutsch U, Assmus W and Lang M 2010 *Eur. Phys. J. B* **77**, 101
[10] Wu M K, Hsu F C, Yeh K W, Huang T W, Luo J Y, Wang M J, Chang H H, Chen T K, Rao S M, Mok B H, Chen C L, Huang Y L, Ke C T, Wu P M, Chang A M, Wu C T and Perng T P 2009 *Physica* C **469** 340
[11] Mizuguchi Y, Tomioka F, Tsuda S, Yamaguchi T and Takano Y 2009 *J. Phys. Soc. Japan* **78** 074712
[12] Fang M H, Pham H M, Qian B, Liu T J, Vehstedt E K, Liu Y, Spinu L and Mao Z Q 2008 *Phys. Rev.* B **78** 224503
[13] Yeh K W, Huang T W, Huang Y, Chen T K, Hsu F C, Wu P M, Lee Y C, Chu Y Y, Chen C L, Luo J Y, Yan D C and Wu M K 2008 *Europhys. Lett.* **84** 37002
[14] Hu R, Bozin E S, Warren J B and Petrovic C 2009 *Phys. Rev.* B **80** 214514

[15] Mizuguchi Y, Tomioka F, Tsuda S, Yamaguchi T and Takano Y 2009 *Appl. Phys. Lett.* **94** 012503
[16] Sales B C, Sefat A S, McGuire M A, Jin R Y, Mandrus D and Mozharivskyj Y 2009 *Phys. Rev.* B **79** 094521
[17] Taen T, Tsuchiya Y, Nakajima Y and Tamegai T 2009 *Phys. Rev.* B **80** 092502
[18] Noji T, Suzuki T, Abe T, Adachi T, Kato M and Koike Y 2010 *J. Phys. Soc. Japan* **79** 084711
[19] Tegel M, Loehnert C and Johrendt D 2010 *Solid State Comm.* **150** 383
[20] Viennois R, Giannini E, van der Marel D, Černy R 2010 *J. Solid State Chem.* **183** 769
[21] Tsurkan V, Deisenhofer J, Günther A, Kant Ch, Krug von Nidda H A, Schrettle F and Loidl A 2010 *arXiv:* 1006.4453 (unpublished)
[22] Rodriguez-Carvajal J 1993 *Physica* B **192** 55
[23] International Tables for Crystallography (Edited by T. Hahn), Kluwer Acad. Publ., Dordrecht, 1996, v. A.
[24] Li S, de la Cruz C, Huang Q, Chen Y, Lynn J W, Hu J, Huang Y L, Hsu F C, Yeh K W, Wu M K and Dai P 2009 *Phys. Rev.* B **79** 054503
[25] Bean C P 1962 *Phys. Rev. Lett.* **8** 250
[26] Bean C P 1964 *Rev. Mod. Phys.* **36** 90
[27] Liu T J, Ke X, Qian B, Hu J, Fobes D, Vehstedt E K, Pham H, Yang J H, Fang M H, Spinu L, Schiffer P, Liu Y and Mao Z Q 2009 *Phys. Rev.* B **79** 174509
[28] Werthamer N R, Helfand E and Hohenberg P C 1966 *Phys. Rev.* **147** 295
[29] Popovich P, Boris A V, Dolgov O V, Golubov A A, Sun D L, Lin C T, Kremer R K and Keimer B 2010 *Phys. Rev. Lett.* **105**, 027003
[30] Braithwaite D, Lapertot G, Knafo W and Sheikin I 2010 *J. Phys. Soc. Japan* **79** 053703
[31] Bouquet F, Wang Y, Fisher R A, Hinks D G, Jorgensen J D, Junod A and Phillips N E 2001 *Europhys. Lett.* **56** 856
[32] Padamsee H, Neighbor J E and Shiffman C A 1973 *J. Low Temp. Phys.* **12** 387
[33] Rotter M, Tegel M, Schellenberg I, Schappacher F M, Pöttgen R, Deisenhofer J, Günther A, Schrettle F, Loidl D and Johrendt D 2009 *New Journal of Physics* **11**, 025014
[34] Kant Ch, Deisenhofer J, Günther A, Schrettle F, Loidl A, Rotter M and Johrendt D 2010 *Phys. Rev.* B **81** 014529
[35] Kato T, Mizuguchi Y, Nakamura H, Machida T, Sakata H and Takano Y. 2009 *Phys. Rev.* B **80** 180507 (R)
[36] Homes C C, Akrap A, Wen J S, Xu Z J, Lin Z W, Li Q and Gu G D 2010 *Phys. Rev.* B **81** 180508
[37] Zhang A M, Xia T L, Kong L R, Xiao J H and Zhang Q M, 2010 J. *Phys. Condens. Matter* **22** 245701
[38] Cheng P, Shen B, Hu J and Wen H H 2010 *Phys. Rev.* B **81** 174529